\documentclass[12pt]{article}
\usepackage[]{epsfig}
\begin{document}
\pagestyle{empty}
     
\begin{center}
\Large\bf
Comment on: ``Search for direct photons from S-Au
Collisions at 200 GeV/nucleon'' by the CERES Collaboration

\vspace{1cm}
\normalsize
{\bf K.H. Kampert$^{1}$, T.C. Awes$^{2}$, P. Stankus$^{2}$,\\
H.H. Gutbrod$^{3}$, V. Manko$^{4}$, T. Peitzmann$^{5}$, and
R. Santo$^{5}$}

\vspace{1cm}
{\small\it
1) Inst. f\"ur exp.\ Kernphysik der Univ.\ Karlsruhe (TH),
     D-76131 Karlsruhe, Germany\\
2) Oak Ridge National Laboratory, Oak Ridge, Tennessee 37831, USA\\
3) SUBATECH, Ecole des Mines de Nantes,F-44070 Nantes, France\\
4) Research Center "Kurchatov Institute", Moscow 123182, Russia\\
5) Inst. f\"ur Kernphysik der Universit\"at M\"unster,
     D-48149 M\"{u}nster, Germany\\
}
\normalsize
\vspace{1cm}
{\small\rm (Submitted to Zeitschrift f\"ur Physik C: 27 Oct.\ 1996;\\
Accepted without modification: 21 March 1997.})
\vspace{1cm}
\end{center}

\begin{abstract}
\noindent In this note we comment on a recent publication in this 
journal by the CERES (NA45) Collaboration \cite{ceres}.  The 
authors report to have determined an upper limit on the direct 
photon yield relative to the decay photon yield in S\,+\,Au 
collisions of 14\,\% and 7\,\% by two different methods, 
respectively.  We argue that these limits are unsupported by the 
results and analysis of the CERES data.  The systematic error 
estimates quoted in the CERES analysis are consistently overly 
optimistic.  Using more realistic estimates of the various error 
contributions and propagating them appropriately we arrive at a 
direct photon upper limit which at best is 20\,\% of the 
inclusive photon yield, and most probably is much higher.
\end{abstract}

\newpage
\pagestyle{plain}
\normalsize

\noindent The authors of Ref.\,\cite{ceres} have presented an 
analysis in which they report to demonstrate an upper limit on 
the production of excess direct photons in 200 AGeV S\,+\,Au 
reactions.  They obtain upper limits, at the 90\,\% confidence 
level, of 14\,\% and 7\,\% of the inclusive photon yield derived 
respectively by two different methods.  This is a remarkable 
achievement by an experiment optimized to measure electron pairs.
Dedicated photon experiments attempting to make similar 
measurements \cite{wa80-photons} have hardly achieved similar 
accuracy.  The reported precision is even more remarkable in 
light of the fact that, as the authors report, only 2.3\% of the 
photons produced are converted and available for measurement, of 
these less than 20\% are identified in the experimental 
apparatus, and of those photons identified, 18\% are background 
photons resulting from Dalitz decays.  Furthermore, in the search 
for excess photons, the measured inclusive yield is to be 
compared to the expected photon yield arising from conventional 
background sources, predominantly radiative decays of $\pi^0$ and 
$\eta$ mesons, none of which have been measured for the CERES 
experimental acceptance and event selection.

Since the low $p_\perp$ direct photon excess is expected to be 
small, if observable at all, in S+Au reactions 
\cite{wa80-photons} it is not surprising that this experiment can 
only set an upper limit.  In setting an upper limit, the crucial 
issue for the experimental measurement is an accurate understanding 
and estimate of the systematic errors.  As the quality and 
importance of the measurement is to be judged by its level of 
accuracy, we consider in detail the analysis of the systematic 
errors presented in Ref.\,\cite{ceres} and we shall demonstrate 
that the authors have decidedly underestimated their systematic 
errors.

Apart from raising concerns about this particular measurement, 
our comments should be viewed in a broader context, since 
preparations are presently being made for the first generation of 
LHC experiments, notably ALICE, where the detection of direct 
photons is one of the aims.  To decide, wether the ``direct'' or 
the conversion method is better suited for this purpose, a 
detailed knowledge of the advantages and disadvantages of the 
different methods is indispensible.  For these reasons we have 
put considerable effort into the comparison of the methods and to 
study the corresponding systematic errors.

In the first method used by the authors an upper limit on the 
integrated direct photon production is extracted by studying the 
ratio $r_\gamma$, given in Eq.(1) of Ref.\,\cite{ceres}, which is 
the ratio of the integrated photon $p_\perp$ distribution, 
integrated from $0.4$ GeV/c $\le p_\perp \le 2.0$ GeV/c, to 
$dN_{ch}/d\eta$.  This ratio is calculated from the experimental 
data, $r_{data}$, and compared to the same ratio calculated from 
simulation for photons arising from hadron decays, $r_{hadr}$.  
The ratio of these two ratios provides a determination of the 
possible direct photon yield in this $p_\perp$ region.  A slight 
direct photon excess of $4\,\%$ is determined in
Ref.\,\cite{ceres}.  The systematic error on these
two ratios is then used to set a direct photon upper limit.

\subsection*{Systematic errors in the measurement
(\boldmath{$r_{data}$})}

\paragraph{Photon reconstruction efficiency:}
A procedure for determination of the photon reconstruction 
efficiency is described: a spectrum of simulated photons is put 
through the simulated detector response and the identified photon 
spectrum is constructed.  The ratio of the reconstructed to 
initial spectrum gives the efficiency.  Obviously, due to the 
finite momentum resolution, the correction function depends on 
assumptions about the original distribution.  Furthermore, due to 
uncertainties in the Cherenkov-Ring reconstruction and 
contaminations by fake rings, it will depend also on the local 
particle density in the apparatus.  Usually, to take such effects 
into account in high precision measurements, an iterative 
procedure for a given particle density distribution is performed 
until the output distribution of the Monte-Carlo (MC) matches the 
observed experimental distribution.  The input distribution of 
the MC is then considered the true one.  However, despite the 
moderate momentum resolution and the strongly varying particle 
density within the angular acceptance window, no similar 
iterative procedure is described in Ref.\,\cite{ceres} and no 
statement is made about how the original distribution was chosen.  
Instead, the authors state qualitatively that the strongly 
$p_{\perp}$ dependent correction function is insensitive to the 
shape of the assumed $p_{\perp}$ input distribution.  Dispite the 
fact that the data are corrected by a correction function which 
varies from a factor of two to a factor of 6 over the range of 
$p_{\perp}$ used, they claim that the yield is determined with an 
extraordinarily small error of only +2.7\,\%, --5\,\%.  Other 
experiments, such as \cite{wa80-photons}, operating at 
90\,\%--100\,\% photon reconstruction efficiency (compared to 
less than 0.5\,\% in this experiment) have achieved just
such a level of precision.  However, in Ref.\,\cite{wa80-photons} 
the accuracy -- after performing such an iterative procedure -- 
is confirmed using the experimental data itself, i.e.\ analyses 
with very different cuts and associated efficiencies result in 
corrected photon distributions with deviations smaller than the 
quoted uncertainty of the reconstruction efficiency.  Since no 
such crucial test is described in Ref.\,\cite{ceres} it is 
difficult to believe such a small error estimate.  A more prudent 
estimate would put a lower limit on the reconstruction efficiency 
of at least $\sigma \approx \pm5\,\%$, and likely greater.

\paragraph{Uncertainty of the momentum scale:}
The momentum scale is said to be known to $\pm$ 2\,\% resulting 
in an error of 2.3\,\% on $r_{data}$. For a purely 
exponential distribution one expects that the uncertainty in the 
integrated photon yield above some lower $p_\perp$ cutoff $p_c$ 
will scale like $e^{\delta p_c/p_0}$, where $\delta$ is the 
percentage uncertainty in the $p_\perp$ slope, $p_0$, or 
equivalently in the momentum scale.  Taking the data of Fig.\,5, 
the inverse exponential slope at $p_{\perp}$ = 0.4 GeV 
corresponds to appr.\ 150 MeV which would then translate into an 
error of $r_{data}$ of 5.5\,\%.  Performing a full power-law fit to 
the data in Fig.\ 5 with a subsequent integration of the yield in 
$0.4 \le p_{\perp} \le 2.0$\,GeV/c yields basically the same 
result, i.e.\ the quoted uncertainty appears to be 
underestimated by approximately a factor of 2.

\paragraph{Conversion probability:}
The uncertainty in the photon convertor thickness enters directly in
the uncertainty in $r_{data}$. The authors suggest incredulously that
the 40 targets were produced independently and therefore that the
uncertainty in the target thickness is only 10\,\%/$\sqrt{40}$ =
1.5\,\%. More likely, the 40 targets were produced in the same
batch from the same foil or measured with the same measuring
device such the errors in the target thicknesses are highly
correlated. Furthermore, the other materials such as the silicon
detectors and windows which contribute 44\,\% of the total conversions
are claimed to have negligible uncertainty in their conversion 
probability. The conversion probability uncertainty is claimed to
be only $\pm3\,\%$, arising solely from the statistics of the Monte
Carlo simulation. Again a more prudent estimate would be that this
uncertainty is at least $\pm5\,\%$ and probably greater.

\paragraph{Uncertainty of $dN_{ch}/d\eta$:}
The authors of Ref.\,\cite{ceres} claim that by measuring the 
total charge on their silicon pad detector which has 64 pads 
covering the region from $1.7 < \eta < 3.4$ they can determine 
the value of $dN_{ch}/d\eta$ at $\eta=2.4$ with a $5\,\%$ 
accuracy.  Considering that this must include corrections for 
$\delta$-electrons, conversions, and slow protons, as well as 
corrections for the varying acceptance with target position and 
the extrapolation from the integrated multiplicity measured to 
the inferred value at $\eta=2.4$ assuming an unmeasured 
$dN_{ch}/d\eta$ distribution, it would seem that the quoted 
uncertainty of only $5\,\%$ is again quite bold.

\subsection*{Systematic errors in the background
calculation (\boldmath{$r_{hadr}$})}
The normalization of the measured photons to the background 
photons expected from conventional sources, most prominently 
$\pi^{0} \to \gamma \gamma$ and $\eta \to \gamma\gamma$ is the 
most critical problem in the data analysis for the reason that none 
of these background sources is measured within the experiment.  As a 
consequence, the authors need to make assumptions about the 
$p_{\perp}$ and rapidity shapes as well as on the total 
integrated yields of these particles.  To accomplish this task, 
the authors rely on transverse momentum spectra and inclusive 
charged particle measurements performed by other experiments for 
different systems, different pseudorapidity windows, and 
different trigger conditions and extrapolate these data, where 
necessary, to the acceptance covered by the CERES experiment.  

Explicitly, the authors start from $p_{\perp}$ and $y$ 
distributions measured by NA34, NA35, WA80, and CERES to 
determine the $\pi^{0}$ \, $p_{\perp}$ distribution.  Each of 
these experiments has made a different measurement; NA34 has 
measured negative pions, NA35 has measured the sum of negative 
hadrons ($\pi^{-} + K^{-} + \bar{p} + \ldots$), WA80 has measured 
neutral pions (i.e.\ those particles which are intended to be 
inferred from this analysis), and CERES has again measured 
charged pions.  The authors then argue to obtain ``reasonable'' 
fits to the available data and arrive, after assumptions about 
particle compositions, different trigger conditions, 
factorization of $d\sigma/dp_{\perp}$ and $y$ (which is known to 
be violated by more than 10\,\%), etc.\ at a systematic
uncertainty of only +3.5\,\%, --5.3\,\% for the photon 
background.  To check these numbers, we have taken tabulated data 
available from each of these experiments and for comparison, 
since only WA80 has measured {\em absolute} cross sections, have 
normalized the $p_{\perp}$ distributions at 0.8\,GeV/c.  Such a 
test exhibits at least an $\approx \pm 25\,\%$ variation of the
cross section ratios in the common $p_{\perp}$ range of the 
experiments.  Because of the unknown absolute cross sections, 
i.e.\ the unknown $p_{\perp}$ value where these spectra should be 
normalized to each other, this error will have to enter directly 
to the background calculation and will obviously be much larger 
than +3.5\,\%, --5.3\,\%.  If the authors instead normalize the
different data sets to the total integrated yields (which method 
was used is not described in the paper) the calculation will 
have to include instead assumptions about an extrapolation of the 
different $p_{\perp}$ distributions to $p_{\perp} = 0$\,GeV/$c$ 
which again will have much larger errors than quoted above.  On 
top of this, in order to put constraints on the photon yield from 
$\pi^{0}$ decay, any procedure based on charged pions
will have to make assumptions about the relative particle 
compositions which will increase the uncertainty further.

Interestingly, the quoted CERES error is similar to or even less 
than the error quoted by WA80 \cite{wa80-photons} for the 
$\pi^{0}$ \, $p_{\perp}$ distribution.  WA80 has measured the 
distribution with the best precision available in nuclear 
reactions and without any assumptions.  Thus, it is questionable 
how utilization of data measured under hardly comparable 
conditions and mostly with very moderate precision combined with 
many assumptions can constrain the $\pi^0$ result as well or 
better than the direct measurement.  We therefore conclude, that 
the authors have underestimated their uncertainty again by at 
least a factor of two.

The actual normalization of the calculated background to the 
shape of the measured photon spectrum in CERES is obtained using 
the $4\pi$ integrated ratio $n_{\pi^{0}}/N_{ch}$.  This quantity 
has been measured with a quoted 5\,\% uncertainty in the forward
hemisphere of 400\,GeV/c pp collisions \cite{aguilar}.  To 
extrapolate this measurement for the purpose of this paper to 
central 200\,AGeV S\,+\,Au reactions, the authors make various 
assumptions about baryon rapidity shifts caused by different 
amounts of nuclear stopping and assume furthermore, that the 
composition of produced particles remains unchanged in the two 
systems.  This scaling assumption is well known to be violated 
since strange particles ($K/\pi$, etc.)  are produced much more 
abundantly in central heavy-ion collisions than in pp collisions.  
Despite these problems, the authors arrive at an overall error of 
only 5.4\,\% (5\,\% of which are from the measurement in
\cite{aguilar} itself), which again appears to be significantly 
underestimated.

To summarize these points, the authors claim to have determined 
the expected number of photons in their acceptance and centrality 
selection with an accuracy which meets or exceeds that obtained 
by other experiments which have actually measured the background 
hadron sources without any assumptions and extrapolations.  This 
has been done using the charged particle multiplicity as the only 
measured quantity.  This is quite a remarkable claim which 
however appears unsubstantiated by a closer look at the 
systematic errors.  

In Table 1 we list the various systematic error contributions 
given in Ref.\,\cite{ceres} for the direct photon upper limit 
estimate.  The error values quoted in Ref.\,\cite{ceres} are 
listed together with values which we would quote being only 
moderately more conservative than in Ref.\,\cite{ceres}.  
Assuming that the various sources of errors are uncorrelated and 
that they may be added quadratically results in a $90\,\%$ CL
direct photon upper limit of at least $20\,\%$. More reasonable
conservative error estimates, as suggested by the above discussion,
would result in a significantly higher upper limit. In this table
we also show that by simply summing the CERES error estimates
quadratically, rather than using their convolution procedure, the
direct photon upper limit is nearly $18\,\%$ rather than $14\%$. 

\bigskip
\begin{table}
\caption{
Summary of various sources of systematic error entering in the 
CERES direct photon upper limit estimate. The CERES error estimates
are shown together with our own estimates. Our total error estimates
have been obtained by summing the individual error contributions 
quadratically. The CERES total error estimate has been obtained
by a more sophisticated analysis.
}\label{errors}
\vspace*{5mm}
\begin{tabular}{llll}
\hline
Quantity: \ \ Source of Error & 
Quoted CERES \hfill & 
Equiv. upper \hfill &
Our estimate \hfill \\
& 
value $(\%)$ \hfill & 
RMS value $(\%)$ \hfill & 
RMS value $(\%)$ \hfill \\
\hline
\hline
$r_{data}$: & & & \\
\ \  $\gamma$ reconstruction efficiency & [-5.,+2.7] & +0.8 & 
+5.0 \ \ (+10.0)$^*$ \\
\ \  Parametrization of efficiency  & $\pm$2.0 & +2.0 & +2.0 \\
\ \  $p_\bot$ uncertainty and integration & $\pm$2.3 & +2.3 & 
+5.0 \\
\ \  $\pi^0$ Dalitz decay contribution & $\pm$5.0 & +5.0 & +5.0 \\
\ \  Conversion probability & $\pm$3.0 & +3.0 & +5.0 \ \ (+10.0)$^*$ \\
\ \  Charged particle density & $\pm$5.0 & +5.0 & +5.0 \\
\hline
Total upper error $\sigma_{r_{data}}$ \hfill & 
\ \ \ \ \ \ \ (+6) & +8.3 \ \ (+9.7) & +11.4 \ (+13.4) \\
\hline
\hline
 $r_{hadr}$: & & & \\
\ \  $\pi^0$ $p_\bot$ shape & [-5.3,+3.5] & +1.0 & +5.0 \ \ (+10.)$^*$ \\
\ \  $\pi^0$ rapidity distribution & [-3.9,+2.4] & +0.7 & +2.4 \\
\ \  $\eta/\pi^0$ scaling & $\pm$4.0 & +4.0 & +4.0 \\
\ \  $n_{\pi^0}/N_{ch}$ normalization & $\pm$5.4 & +5.4 & 
+5.4 \ \ (+10.)$^*$ \\
\hline
Total upper error $\sigma_{r_{hadr}}$ \hfill &
\ \ \ \ \ \ \ (+7) & +6.8 \ \ (+8.0) & +8.7 \ \ (+10.2)  \\
\hline
\hline
Total upper error $\sigma_{r_{data}/r_{hadr}}$ \hfill & 
\ \ \ \ \ \ \ (+9) & +10.7 \ (+12.6) & +14.3 \ (+16.9) \\
\hline
\hline
\\
\multispan3{ $90\%$ CL upper limit on direct photons \hfill}  &  \\
($4\% + 1.28\cdot \sigma_{r_{data}/r{hadr}}$) \hfill &
$14\%$ & $17.7\%$ & $20.1\%$ \\
\hline \\
\multispan4{Quantities in () on total errors indicate FWHM/2 values, as quoted
in Ref.\,\cite{ceres}. \hfill} \\
\multispan3{$^*$ Reasonable conservative estimate (not
used  here). \hfill} \\

\end{tabular}
\end{table}
\vskip -0.3cm

\subsubsection*{Multiplicity Dependence of the Photon Production}

The second method of analysis described in Section 5 of 
Ref.\,\cite{ceres} attempts to extract information about the 
production of direct photons through the multiplicity dependence 
of the inclusive photon yield.  In this method an upper limit is 
obtained on the coefficient $\alpha$ of a possible term which is 
quadratic in the reconstructed $dN_{ch}/d\eta$.  This is an 
interesting approach which avoids the above noted problems of 
absolute normalization.  However, it must be emphasized that such 
a measurement provides at best only indirect information about a 
possible direct photon excess.  On general theoretical grounds it 
might be expected that the direct photon production should scale 
like the square of the local charge density in the hot dense 
medium, but it is obvious that the observed variation in the 
reconstructed $dN_{ch}/d\eta$ is not directly related to a change 
in the local charge density.  That is, it is extremely unlikely 
that the observed factor of nearly four increase in 
$dN_{ch}/d\eta$ corresponds to a factor of four increase in local 
density and hence an expected factor of 16 increase in direct 
photon production.  Thus, extracting an upper limit on the 
quadratic dependence of the total photon production on the 
reconstructed $dN_{ch}/d\eta$ cannot be said to correspond to an 
upper limit on the direct photon production without some 
estimation of this dilution factor, which is most probably a very 
large factor.  Other effects such as resolution of the 
multiplicity measurement will further dilute any such 
correlation.  Therefore, to discuss upper limits on the quadratic 
coefficient in the same context as a discussion of the direct 
photon upper limit without clarification of their weak 
relationship is extremely misleading.  A direct photon upper 
limit is probably many times greater than the determined upper 
limit on the quadratic dependence.

Furthermore, what is determined in Ref.\,\cite{ceres} is an upper 
limit on the quadratic coefficient, $\alpha$.  It cannot be 
translated into an upper limit on the percentage of direct 
photons having a quadratic dependence on $dN_{ch}/d\eta$ without 
specifying a value for $dN_{ch}/d\eta$.  Therefore it is again 
misleading to quote a ''limit of $7\,\%$ ($90\,\%$ CL) for the 
strength of a photon source with a quadratic multiplicity 
dependence.''  as in the conclusion and abstract of the paper 
without stating that this corresponds to the value at 
$dN_{ch}/d\eta = 131$.  Quoted in these terms the limit could be 
reduced arbitrarily by chosing a lower multiplicity, or would 
increase by nearly a factor of two for the highest multiplicities 
measured.

Although discussed in much less detail than for the first 
analysis method, the errors quoted for this analysis again appear 
to be severely underestimated.  It is curious that the total 
systematic errors in some cases seem to be less than or equal to 
some of the individual error contributions.  Furthermore, simply 
accepting the quoted final errors one calculates an upper limit 
which is more than $20\,\%$ greater than the upper
limit which is quoted. Finally, we believe that there is a 
major source of uncertainty which is not discussed but which brings
into question the results and conclusions of this entire analysis.

\begin{figure}[p]
   \centerline{\epsfxsize=11cm \epsfbox{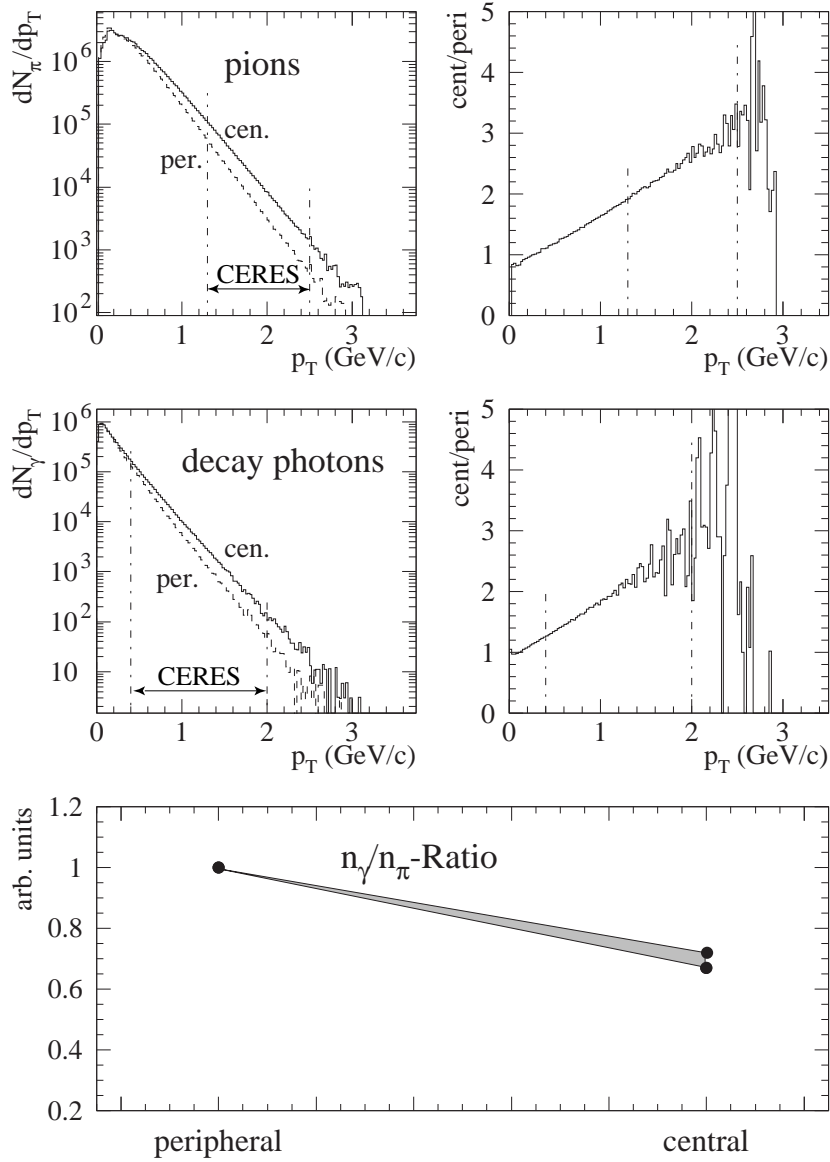}}
   \caption{Simulation results to estimate the effect of changes 
   in the spectral shape to $n_{\gamma}/N_{\pi}$. The two upper 
   left histograms show the $\pi^{0}$ and decay photon spectra, 
   respectively. Full lines are for central events, and dashed 
   lines for peripheral events. The integration intervals used by 
   CERES are indicated. The upper right spectra show the ratios 
   of the individual distributions and demonstrate that the 
   amount of change between central and peripheral events is 
   taken conservatively. The bottom figure shows the expected 
   drop in the observed $n_{\gamma}/N_{\pi}$ ratio. The upper 
   bound is for exponential $\pi^0$ distributions with slope 
   parameters as given in the text, and the lower point is for a 
   power law spectrum describing the experimental $\pi^{0}$ WA80 
   data.}
   \protect\label{ratio}
\end{figure}

The problem is twofold: ({\em i}) if true at all, the assumption 
of a squared multiplicity scaling of photons over neutral pions 
holds for the full phase space integrated yields only, whereas in 
this experiment charged pions and photons are reconstructed only 
in a limited rapidity range and at rather large transverse 
momenta.  ({\em ii}) $dN/dy$ as well as the $p_{\perp}$ spectra 
of pions and photons are known to depend themselves on 
centrality, i.e.\  on the multiplicity of particles.  Here, we 
shall only estimate the effect caused by the change in the 
transverse momentum spectra when going from peripheral to central 
collisions.  For simplicity, we will again start with exponential 
distributions for which the numbers given below are easy to 
verify.  Assuming an inverse slope parameter of pions 
$1/p_{\perp} dN/dp_{\perp} \propto \exp(-p_{\perp}/p_{0})$ with 
$p_{0}$ increasing from 200 to 220 MeV when going from peripheral 
to central collisions \cite{santo}, the fraction of observed 
pions in the $p_{\perp}$ range from 1.3 -- 2.5 GeV/$c$ will 
increase by 66\,\%.  The decay photons, on the other hand, can be 
rather well described by $dN/dp_{\perp} \propto 
\exp(-p_{\perp}/p_{0})$, with $p_{0}$ increasing from 190 to 210 
MeV/$c$.  The yield observed in $0.4 \le p_{\perp} \le 
2.0$~GeV/$c$ will then increase by 22\,\%, so that one expects 
$n_{\gamma}/N_{\pi}$ in the acceptance of CERES to drop by more 
than 25\,\%.  In a more detailed analysis we have verified this 
analytical estimate by performing a full Monte-Carlo simulation 
of the $\pi^{0}$ decay, including the photons from $\eta$ decay 
\cite{wa80eta}, and taking into account the CERES experimental 
acceptance.  The result is basically unchanged, i.e.\ we find a 
drop of $n_{\gamma}/N_{\pi}$ by approx.\ 25\,\%.  Furthermore
this might be considered a lower bound on the expected variation 
since pure exponential distributions have been assumed down to 
very low $p_{T}$.  The assumption of power-law $\pi^{0}$ \, 
$p_{\perp}$ spectra as seen by WA80 and other experiments for 
peripheral and central reactions \cite{santo}, results in a drop 
by approx.\ 33\,\%.  These calculations and their results are 
illustrated in Fig.\,1, where the original $\pi^{0}$ spectra 
(power law spectral shape in $d\sigma/dp_{T}^{2}$) as well as 
their photon-decay spectra are plotted.  Since a drop as seen in 
the bottom graph would be easily visible in Fig.\,6 of 
Ref.\,\cite{ceres}, one is puzzled to understand what might have 
compensated for it.

To summarize this point, any measurement based on integrating 
photons or pions, in a range whose lower limit in $p_{\perp}$ is 
well above $m_{\pi}$ will be {\em inherently} unable to 
distinguish between a change in overall number of particles {\em 
vs} a change in the slope/shape of the distribution down to very 
low $p_{\perp}$.  The experiment is unable to control the 
systematic variations of $n_{\gamma}/N_{\pi}$ which are of the 
order of 25\,\% or more.  Nevertheless the authors claim to have 
measured this quantity to 5\,\% accurate.

\subsection*{Summary and Conclusion}

In conclusion, the major task of an experiment presuming to 
present an upper limit measurement is to make a reliable 
assessment of the errors of the measurement.  Setting an upper 
limit on direct photon production in the CERES experiment is a 
formidable task.  The experiment measures only less than 0.5\,\% 
of the photons in their acceptance.  From this they must subtract 
the photon background contribution arising dominantly from the 
decay of $\pi^{0}$ and $\eta$ mesons.  These dominant background 
photon sources are not measured within CERES. Yet a 90\,\% CL 
direct photon limit of 14\,\% or 7\,\%, depending on method of 
analysis, is claimed.  A critical reader cannot help but be 
skeptical.  The analysis presented in Ref.\,\cite{ceres} 
generally does not present results to support their error 
estimates.  According to our own estimates and investigations 
most of the error estimates presented in Ref.\,\cite{ceres} are 
overly optimistic, as we have indicated in this comment.  More 
careful consideration of the possible errors in the extraction of 
the integrated direct photon yield would indicate a 90\,\% CL 
upper limit of no lower than 20\,\%, and most likely much higher.  
The method of analysis of the quadratic dependence of the photon 
yield on $dN_{ch}/d\eta$ has large systematic uncertainties which 
have not been adequately addressed in Ref.\,\cite{ceres}.  With 
the present understanding, the limits on the quadratic dependence 
cannot be used to infer anything of significance on the direct 
photon production.


\end{document}